\documentclass[twocolumn,showpacs,preprintnumbers,fleqn,amsmath,amssymb,prl]{revtex4}
\usepackage{graphicx}
\usepackage{time}
\usepackage{dcolumn}
\usepackage{bm}

\begin{document}
\title{Giant planar Hall effect in colossal magnetoresistive \textbf{${\rm La_{0.84}Sr_{0.16}MnO_{3}}$} thin films}
\author{Y. Bason}
\author{L. Klein}
\affiliation{Department of Physics, Bar Ilan University, Ramat Gan
52900, Israel}

\author{J.-B. Yau}
\author{X. Hong}
\author{C. H. Ahn}
\affiliation{Department of Applied Physics, Yale University, New
Haven, Connecticut 06520-8284, USA}
\date{\today}

\begin{abstract}
The transverse resistivity in thin films of ${\rm
La_{0.84}Sr_{0.16}MnO_3}$ (LSMO) exhibits sharp field-symmetric
jumps below ${\rm T_c}$. We show that a likely source of this
behavior is the giant planar Hall effect (GPHE) combined with
biaxial magnetic anisotropy. The effect is comparable in magnitude
to that observed recently in the magnetic semiconductor Ga(Mn)As.
It can be potentially used in applications such as magnetic
sensors and non-volatile memory devices.
\end{abstract}
\maketitle

The planar Hall effect (PHE) \cite{phe1} in magnetic conductors
occurs when the  resistivity depends on the angle between the
current density \textbf{J} and the magnetization \textbf{M}, an
effect known as anisotropic magnetoresistance (AMR) \cite{amr1}.
When $\textbf{M}$ makes an angle $\theta$ with $\textbf{J}$, the
AMR effect is described by the expression
$\rho=\rho_{\perp}+(\rho_{\parallel}-\rho_{\perp})\cos^2\theta$,
where $\rho_{\perp}$ and $\rho_{\parallel}$ are the resistivities
for \textbf{J $\perp$ M} and \textbf{J $\parallel$ M},
respectively.  The AMR yields a transverse ``Hall-like'' field if
\textbf{J} is not parallel or perpendicular to $\textbf{M}$.
Assuming $\textbf{J}=J_x\hat{x}$ and $\textbf{M}$ are in the $x-y$
plane with an angle $\theta$ between them, the generated electric
field has both a longitudinal component:

\begin{equation}
\\E_x=\rho_{\perp}j_x+(\rho_{\parallel}-\rho_{\perp})j_x \cos^2 \theta,
\label{par}
\end{equation}
and a transverse component:
\begin{equation}
\\E_y=(\rho_{\parallel}-\rho_{\perp})j_x \sin \theta \cos \theta.
\label{per}
\end{equation}
This latter component is denoted the planar Hall effect.  Unlike
the ordinary and extraordinary Hall effects, the PHE shows an even
response upon inversion of $\textbf{B}$ and $\textbf{M}$;
therefore, the PHE is most noticeable when $\textbf{M}$ changes
its axis of orientation, in particular between $\theta=45^\circ$
and $\theta=135^\circ$.

The PHE in magnetic materials has been previously investigated in
3$d$ ferromagnetic metals, such as Fe, Co and Ni films, as a tool
to study in-plane magnetization \cite{ipm1}.  It has also been
studied for low-field magnetic sensor applications \cite{phs}.
Recently, large resistance jumps in the PHE have been discovered
in the magnetic semiconductor Ga(Mn)As below its Curie
temperature, $\sim 50 \ {\rm K}$ \cite{gphe1}.  Four orders of
magnitude larger than what has been observed in ferromagnetic
metals, it is called the giant planar Hall effect (GPHE). Ga(Mn)As
exhibits biaxial magnetocrystalline anisotropy; consequently, the
magnetization reversal in a field scan occurs in two steps of
$90^\circ$ rotations. When the current path lies between the two
easy axes, the $90^\circ$ rotations lead to switching-like
behavior in the PHE, which is similar to the switching resistivity
curves observed in giant magnetoresistance (GMR) heterostructures
\cite{gmr1} and tunneling magnetoresistance (TMR) trilayers
\cite{tmr1}. This suggests that the GPHE in magnetic materials may
be suitable for applications in spintronics \cite{Spintronics},
such as field sensors and non-volatile memory elements.

\begin{figure}
\begin{center}
\includegraphics [scale=0.4,trim=0 0 0 0]{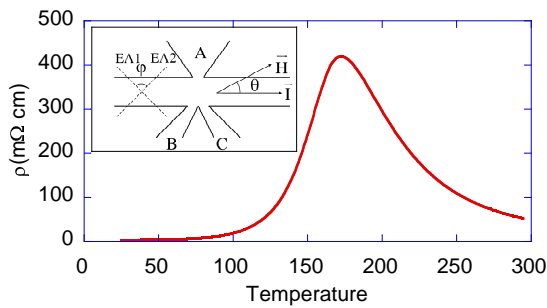}
\end{center}
\caption{$\rho$ vs. T for an LSMO thin film. Inset: The pattern
used for resistivity and Hall measurements. The two easy axes
directions (EA1 and EA2) and the angle ($\theta$) between the
applied field and the current are also shown. The current path is
along either the [100] or  [010] direction.} \label{sketch}
\end{figure}

Here we report on the GPHE observed in the colossal
magnetoresistive material (CMR), $\rm{La_{1-x}Sr_{x}MnO_3}$
(LSMO). When $x$ is between 0.15 and 0.3, LSMO is a ferromagnetic
metal at low temperatures and a paramagnetic insulator at high
temperatures, with the Curie temperature coinciding with the
metal-insulator transition temperature.  Depending on the carrier
concentration, the Curie temperature of LSMO ranges from 150 K to
350 K. Here, we report on films with a doping level of ${x} \sim
0.16$ and resistivity-peak temperature of $\rm{\sim 180\ K}$ (see
Fig. \ref{sketch}). The films exhibit transverse resistivity jumps
comparable to that observed in Ga(Mn)As, and they persist up to
temperatures $>$ 140 K.

Thin films (about 40 nm) of LSMO have been deposited epitaxially
on single-crystal [001] ${\rm SrTiO_3}$ substrates using off-axis
magnetron sputtering.  $\theta - 2\theta$ x-ray diffraction
reveals c-axis oriented growth (in the pseudo-cubic frame), with a
lattice constant of ${\rm \sim 0.385 \ nm}$, consistent with a
strained film \cite{constrained}. No impurity phases are detected.
Rocking curves taken around the 001 reflection have a typical full
width at half maximum of $0.05^\circ$.  The film surface has been
characterized using atomic force microscopy (AFM), which shows a
typical root-mean-square surface roughness of ${\rm \sim 0.2 \
nm}$.  The films are patterned into Hall bars using
photolithography for longitudinal and transverse resistivity
measurements (see Fig. \ref{sketch}), with current paths along the
[100] and [010] directions.

\begin{figure}
\begin{center}
\includegraphics[scale=0.4,trim=0 0 0 0]{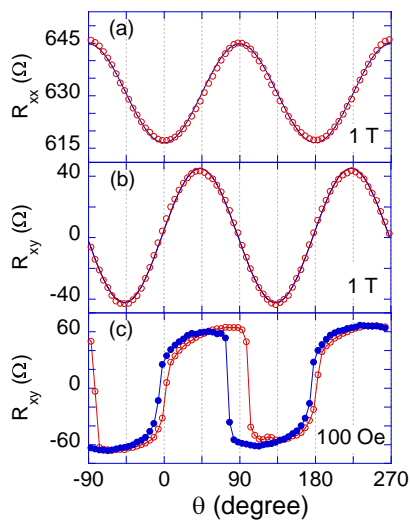}
\end{center}
\caption{Measurements of $R_{xx}$ and $R_{xy}$ vs. $\theta$ at
$T=120 \ {\rm K}$. (a) $R_{xx}$ measured between B and C. The line
is a fit to $\cos^2\theta$. (b) $R_{xy}$ measured between A and C.
The line is a fit to $\sin\theta \cos\theta$. (c) $R_{xy}$
measured between A and C with H=100 Oe.} \label{layout}
\end{figure}

We first investigate the AMR in the LSMO films with a constant
magnetic field applied in the plane of the film. Figure
$\ref{layout}$ shows the transverse resistivity and the
longitudinal resistivity as a function of $\theta$, the angle
between the applied magnetic field and the current. The
longitudinal resistance, $R_{xx}$, is measured between B and C
(see Fig. $\ref{sketch}$). The transverse resistance, $R_{xy}$, is
obtained by measuring the resistance between A and C and
subtracting the longitudinal component based on the $R_{xx}$
measurement.  At high fields the magnetization is expected to be
parallel to the applied field.  We find that $R_{xx}(\theta)$ has
a $\cos^2 \theta$ dependence while $R_{xy}(\theta)$ has a $\sin
\theta \cos \theta$ dependence. At lower fields, the angular
dependence changes, as the effect of the magnetocrystalline
anisotropy becomes significant, and we observe sharp switches in
the PHE (see Fig. $\ref{layout}$c). We interpret the switches as
jumps between easy axes; since the symmetry axes for the
switchings are $\theta=0^\circ$ and $\theta= 90^\circ$ it is
reasonable that the easy axes are in between, namely at
$\theta=45^\circ$ and $\theta= 135^\circ$.

\begin{figure}
\begin{center}
\includegraphics[scale=0.4,trim=0 0 0 0]{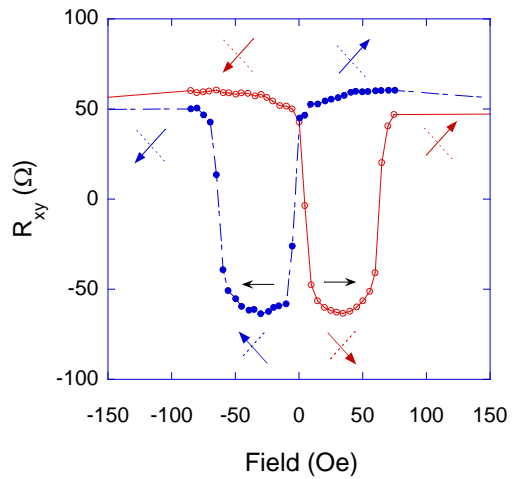}
\end{center}
\caption{PHE vs. H at 120 K with $\theta=10^\circ$. The arrow
shows the magnetization direction along one of the easy axes while
the dashed lines indicate the other easy axis direction. The
horizontal arrows indicate the field sweep directions.}
\label{bigsw}
\end{figure}

Figure \ref{bigsw} shows the switching behavior as a function of
field sweeps with $\theta=10^\circ$. At high positive field, the
magnetization is parallel to the applied field, and the PHE is
positive.  As the field is reduced, the magnetization gradually
aligns along the easy axis closer to the field orientation (EA2).
As the field orientation is reversed, the magnetization first
switches to the other easy axis (EA1), which is an intermediate
state with a negative PHE. As the field becomes more negative, the
magnetization goes back to the initial easy axis (EA2), but with
opposite polarity. A similar process happens when the field is
scanned from negative to positive field.

\begin{figure}
\begin{center}
\includegraphics[scale=0.4,trim=00 0 0 0]{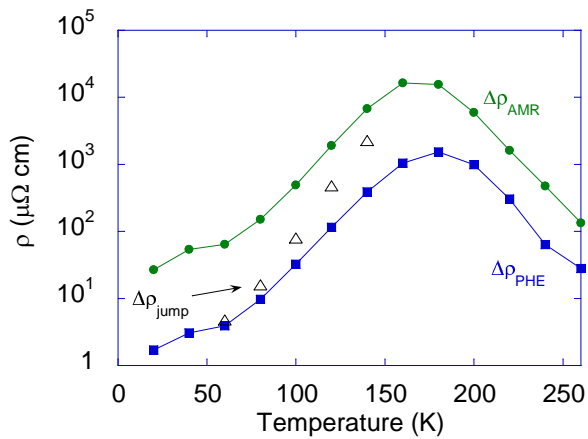}
\end{center}
\caption{ $\Delta \rho_{AMR}$ (connected circles), $\Delta
\rho_{PHE}$ (connected squares) - both measured in a 4 T field,
and $\Delta \rho_{jump}$ (unconnected triangles) vs. T.
$\Delta\rho_{jump}$ is extracted at lower fields. The lines are
guide to the eye.} \label{compare}
\end{figure}

The temperature dependence of the switching shows that the jumps
decrease rapidly as a function of temperature (Fig.
\ref{compare}). Based on the fits to the experimental data (as
presented in Fig. $\ref{layout}$) and Eqs. \ref{par} and
\ref{per}, we calculate $\Delta \rho= \rho_\parallel - \rho_\perp$
at different temperatures. Figure \ref{compare} shows $\Delta\rho$
extracted from the AMR ($\Delta\rho_{AMR}$), the PHE
($\Delta\rho_{PHE}$) and the field sweep jump measurements
($\Delta \rho_{jump}$) as a function of temperature. An in-plane
magnetic field of 4 T was used to extract $\Delta \rho_{AMR}$ and
$\Delta\rho_{PHE}$ at all temperatures. We see that
$\Delta\rho_{AMR}$ and $\Delta\rho_{PHE}$ show similar temperature
dependencies; however, there is a significant difference in their
magnitude \cite{variation}. Considering possible sources for this
difference, we note that Eqs. \ref{par} and \ref{per} are based on
the assumption of uniform current, while the manganites are
intrinsically inhomogeneous and exhibit percolative current paths
\cite{manganites-rev}. In addition, these equations are expected
to be valid for an isotropic medium. Here, the films are epitaxial
and the role of crystal anisotropy is yet to be determined.

As shown in Fig. \ref{compare}, the AMR and GPHE are also observed
above $\rm{T_c}$, and while switching is naturally not observed,
the GPHE may still be interesting for applications where
non-hysteretic behavior in field is required, such as Hall
sensors.

Bi-axial magnetic anisotropy in (001) LSMO films has previously
been reported \cite{suzuki}, and there have been studies of
biaxial anisotropy \cite{lcmo1} and AMR \cite{lcmo2} in other
colossal magnetoresistance materials, such as ${\rm
La_{1-x}Ca_xMnO_3}$ (LCMO). Therefore, one may expect to observe
the GPHE and switching behavior in CMR materials with other doping
levels and chemical compositions.

In conclusion, we have observed the GPHE in LSMO thin films at
temperatures as high as 140 K. By optimizing the chemical
composition and the device geometry, one may expect a larger
effect at higher temperatures, thus allowing for the application
of the GPHE in manganites, such as magnetic sensors and
non-volatile memory devices.

\begin{acknowledgments}
L. K. and C. H. A. acknowledge support from Grant No. 2002384 from
the United States - Israel Binational Science Foundation (BSF),
Jerusalem, Israel.
L.K. acknowledges support by the Israel Science
Foundation founded by the Israel Academy of Sciences and
Humanities.  Work at Yale supported by AFOSR and NSF.
\end{acknowledgments}


\begin{thebibliography}{999}
\bibitem{variation} The difference in the magnitude of $\Delta\rho_{AMR}$ and $\Delta\rho_{PHE}$ is also sample dependent.
\end{thebibliography}
\end{document}